\documentclass[prb,showpacs,amsfonts,amsmath,twocolumn,superscriptaddress,a4paper]{revtex4-1}
\usepackage{array,amsmath,amsfonts,amssymb,dsfont}
\usepackage{natbib}
\usepackage[T1]{fontenc}
\usepackage{ae}
\usepackage{color}
\usepackage{graphicx}

\begin{document}

\title{Geometric phases in superconducting qubits beyond the two-level-approximation}

\author{S.~Berger}
 \email{sberger@phys.ethz.ch}
 \affiliation{Department of Physics, ETH Zurich, CH-8093 Zurich, Switzerland}
\author{M.~Pechal}
 \affiliation{Department of Physics, ETH Zurich, CH-8093 Zurich, Switzerland}
\author{S.~Pugnetti}
 \affiliation{NEST, Scuola Normale Superiore and Istituto Nanoscienze -- CNR, 56126 Pisa, Italy}
\author{A.~A.~Abdumalikov~Jr.}
 \affiliation{Department of Physics, ETH Zurich, CH-8093 Zurich, Switzerland}
\author{L.~Steffen}
 \affiliation{Department of Physics, ETH Zurich, CH-8093 Zurich, Switzerland}
\author{A.~Fedorov}
 \affiliation{Department of Physics, ETH Zurich, CH-8093 Zurich, Switzerland}
\author{A.~Wallraff}
 \affiliation{Department of Physics, ETH Zurich, CH-8093 Zurich, Switzerland}
\author{S.~Filipp}
 \affiliation{Department of Physics, ETH Zurich, CH-8093 Zurich, Switzerland}


\renewcommand{\i}{{\mathrm i}}
\def\1{\mathchoice{\rm 1\mskip-4.2mu l}{\rm 1\mskip-4.2mu l}{\rm 1\mskip-4.6mu l}{\rm 1\mskip-5.2mu l}}
\newcommand{\ket}[1]{|#1\rangle}
\newcommand{\bra}[1]{\langle #1|}
\newcommand{\eval}[1]{\langle #1\rangle}
\newcommand{\braket}[2]{\langle #1|#2\rangle}
\newcommand{\ketbra}[2]{|#1\rangle\langle#2|}
\newcommand{\opelem}[3]{\langle #1|#2|#3\rangle}
\newcommand{\projection}[1]{|#1\rangle\langle#1|}
\newcommand{\scalar}[1]{\langle #1|#1\rangle}
\newcommand{\op}[1]{\hat{#1}}
\newcommand{\vect}[1]{\boldsymbol{#1}}
\newcommand{\id}{\text{id}}
\newcommand{\red}[1]{\textcolor{red}{#1} }

\newcommand{\rad}{\ensuremath{\, \mathrm{rad}}}
\newcommand{\khz}{\ensuremath{\, \mathrm{kHz}}}
\newcommand{\mhz}{\ensuremath{\, \mathrm{MHz}}}
\newcommand{\ghz}{\ensuremath{\, \mathrm{GHz}}}
\newcommand{\dbm}{\ensuremath{\, \mathrm{dBm}}}
\newcommand{\us}{\ensuremath{\, \mathrm{\mu s}}}
\newcommand{\ns}{\ensuremath{\, \mathrm{ns}}}
\newcommand{\mk}{\ensuremath{\, \mathrm{mK}}}
\newcommand{\um}{\ensuremath{\, \mathrm{\mu m}}}
\newcommand{\nm}{\ensuremath{\, \mathrm{nm}}}

\begin{abstract}
Geometric phases, which accompany the evolution of a quantum system and depend only on its trajectory in state space, are commonly studied in two-level systems. Here, however, we study the adiabatic geometric phase in a weakly anharmonic and strongly driven multi-level system, realised as a superconducting transmon-type circuit.
We measure the contribution of the second excited state to the two-level geometric phase and find good agreement with theory treating higher energy levels perturbatively. By changing the evolution time, we confirm the independence of the geometric phase of time and explore the validity of the adiabatic approximation at the transition to the non-adiabatic regime.
\end{abstract}

\maketitle

When a quantum mechanical system evolves under a time-dependent Hamiltonian, its wavefunction acquires a geometric phase in addition to the dynamic phase.\cite{wilczek} If the evolution of the Hamiltonian is cyclic and adiabatic, the geometric phase is termed Berry's phase.\cite{Berry84} While the dynamic phase is proportional to the time integral of the system energy, Berry's phase simply depends on the geometry of the path the Hamiltonian traces out in its parameter space and is independent of time and energy.  It has been observed in a variety of systems, ranging from photons in optical fibers~\cite{Tomita1986} and nuclear magnetic resonance (NMR)~\cite{Jones2000} to superconducting circuits~\cite{Leek2007,Neeley2009}, superconducting charge pumps\cite{Mottonen2008} and electronic harmonic oscillators.\cite{Pechal2012}
The purported resilience of the geometric phase against certain types of noise, studied both in theory~\cite{De2003,Carollo2003a,Whitney2005} and experiment~\cite{Leek2007,Filipp2009a,Cucchietti2010} has raised interest for implementing geometric gates in quantum information processing.
Recently, the geometric aspects of multi-level systems have attracted increased attention. The geometric phase has been observed in NMR interferometry in a three-level system.\cite{Chen2009a} A superconducting phase qudit has been employed as an effective four-level system to show the symmetry of spinors under $2\pi$-rotations, which can be interpreted as a geometric phase.\cite{Neeley2009} It has also been proposed to use three-level systems to detect non-Abelian geometric phases~\cite{Du2011} and schemes to perform non-adiabatic holonomic quantum computation have been studied theoretically.\cite{Sjoqvist2011}

\begin{figure}
  \centering
  \includegraphics[width=86mm]{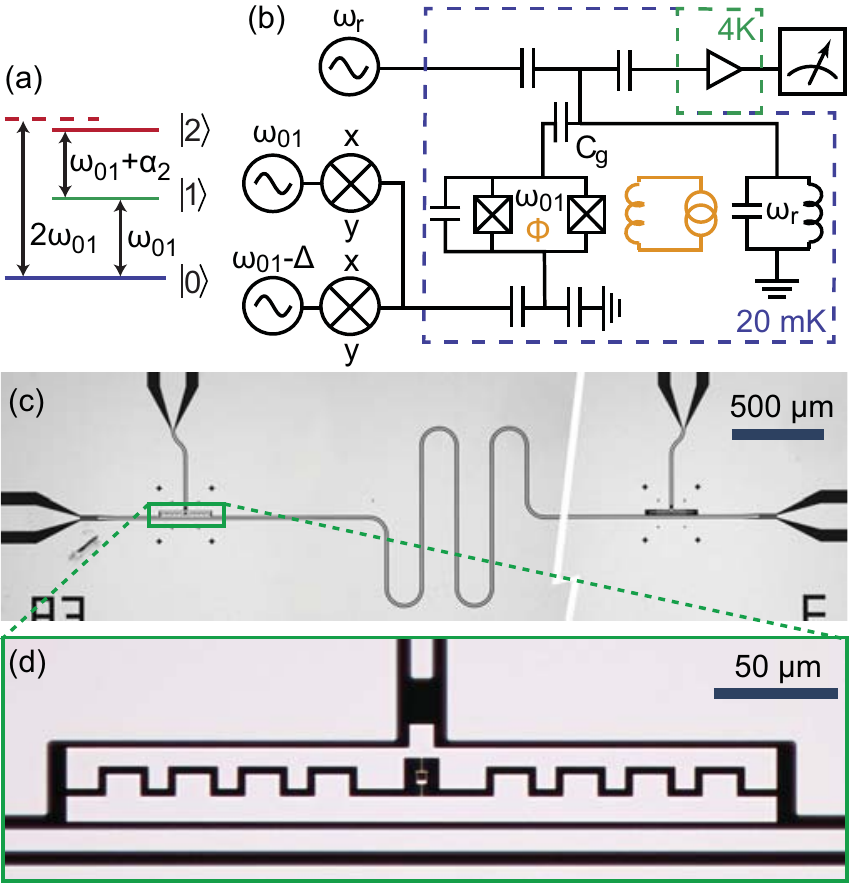}
  \caption{(Color online)
  (a) Schematic energy level diagram of the transmon, with the computational subspace spanned by $\ket{0}$ and $\ket{1}$, and the second excited state $\ket{2}$ with anharmonicity $\alpha_2<0$.
  (b) Lumped element circuit diagram of the sample and the measurement setup (see text for details).
  (c) Optical microscope image of the sample with two transmons coupled to a coplanar waveguide resonator with individual capacitively coupled microwave gate lines.
  (d) Close-up of the transmon used in the experiments.
  }
  \label{fig:sample}
\end{figure}

Here, we present measurements of the effect of higher energy levels on the controlled accumulation of geometric phases in a superconducting qubit system.
In contrast to previous measurements of the geometric phase in superconducting circuits, where a Cooper-pair-box (CPB) qubit was used,\cite{Leek2007} we employ a transmon-type qubit.\cite{Koch2007}
It is characterised by an increased ratio $E_J/E_C$ of Josephson energy to charging energy of the qubit, a smaller charge dispersion~\cite{schreier2008} and a reduced anharmonicity, typically only a few hundred MHz.
As a consequence, the $\ket{1}\leftrightarrow\ket{2}$ transition of frequency $\omega_{12}$ is in close vicinity to the $\ket{0}\leftrightarrow\ket{1}$ transition of frequency $\omega_{01}$ [Fig.~\ref{fig:sample}(a)].
Therefore, whereas the CPB qubit can safely be approximated as a two-level system, the transmon cannot. Higher transmon levels affect the qubit dynamics,\cite{Motzoi2009,Chow2010} but can also be employed as a resource for quantum gates,\cite{DiCarlo2009} to improve single-shot readout,\cite{Mallet2009} or to implement a single-photon router.\cite{Hoi2011} Quantum optical experiments involving three levels have been carried out~\cite{Baur2009,Sillanpaa2009,Abdumalikov2010} and their controlled preparation and tomography has been demonstrated.\cite{Bianchetti2010}

Here, we find that the difference in the level structure between the transmon and an effective two-level system also significantly affects the geometric phase. Time-independent perturbation theory is used to successfully model the geometric phase of a transmon. Then, the dependence of the geometric phase on evolution time in both the adiabatic and non-adiabatic regime is analysed. Non-adiabatic corrections to the geometric phase are observed and explained by simulating unitary dynamics.

For manipulation and readout, the transmon is coupled to a coplanar transmission line resonator of quality factor $Q=2155$ via a gate capacitance $C_{\mathrm{g}}$. The lumped-element circuit diagram of the measurement setup, as well as an optical microscope image of the sample, is shown in Fig.~\ref{fig:sample}(b-d). The sample is mounted in a dilution refrigerator and operated at a temperature of $20\mk$. The transition frequency of the transmon is tunable by externally applied magnetic flux $\Phi$, generated using superconducting coils mounted underneath the sample. The transmon state is manipulated using resonant and off-resonant microwave tones (of frequency $\omega_{01}$ and $\omega_{01}-\Delta$, respectively) created by AC modulation of an in-phase/quadrature-mixer. This provides individual control of both quadratures ($x$ and $y$) of the driving microwave signal, which couples capacitively to the transmon via a local microwave gate line.

\begin{figure}
  \centering
  \includegraphics[width=86mm]{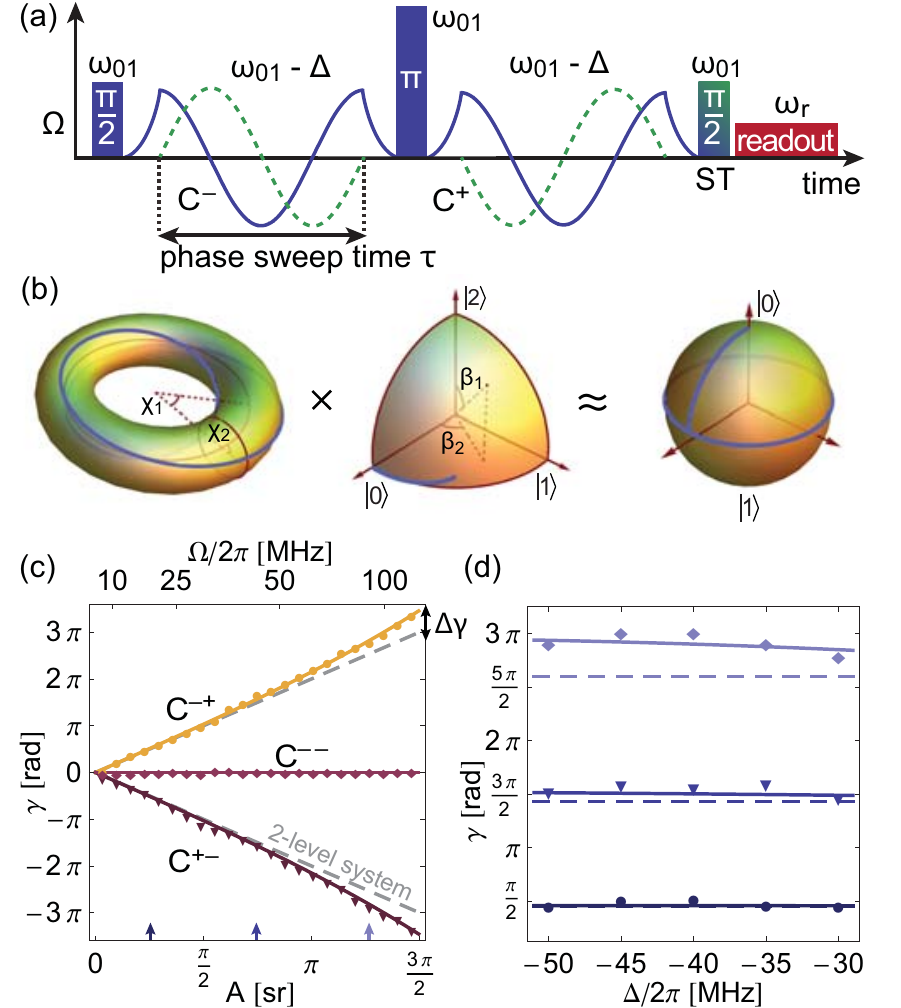}
  \caption{(Color online)
  (a) Sketch of the microwave pulse sequence used in a geometric phase measurement, consisting of a series of resonant ($\omega_{01}$) and off-resonant ($\omega_{01}-\Delta$) pulses applied either on the x-quadrature (blue solid line) or the y-quadrature (green dotted line). The geometric phase is generated by the adiabatic evolution of the qubit between the resonant pulses.
  (b) Simulated adiabatic evolution (blue line) of the ground state subjected to the off-resonant drive along the path $C^-$, visualised in the Hilbert space of a three-level system and on the Bloch sphere, the approximate two-level equivalent (see text for details).
  (c) Extracted phase $\gamma$ as a function of solid angle $A$. Shown is the experimental data for $C^{-+}$ (circles), $C^{--}$ (diamonds) and $C^{+-}$ (triangles), as well as the geometric phase obtained with second-order perturbation theory (solid lines) and the prediction for a two-level system (dashed lines). The off-resonant pulses were applied with detuning $\Delta/2\pi=-35\mhz$.
  (d) Extracted phase $\gamma$ as a function of detuning $\Delta$. The experimental data for solid angles $A\approx\pi/4,3\pi/4,$ and $5\pi/4$ [indicated by circles, triangles and diamonds, respectively, and also indicated by arrows in (c)] is shown alongside the geometric phase calculated using second-order perturbation theory (solid lines) and the prediction for a two-level system (dashed lines).}
  \label{fig:phase}
\end{figure}

From spectroscopic measurements, we have determined a maximum Josephson energy $E_J/h=13.96\ghz$ and a charging energy $E_C/h=0.36\ghz$, which corresponds to a
maximum transition frequency $\omega_{01,\mathrm{max}}/2\pi=5.95\ghz$, and a coupling strength $g/2\pi=360\mhz$ to the fundamental mode of the resonator. To reduce the Purcell effect and optimise coherence properties, $\omega_{01,\mathrm{max}}$ was designed to lie below the fundamental mode of the resonator.\cite{Houck2008} The experiment is carried out in the dispersive regime, where the transmon, biased at $\omega_{01,\mathrm{max}}$, is non-resonantly coupled to the resonator (at frequency $\omega_r/2\pi=6.662\ghz$ with the qubit in the ground state) and can be read out via a state-dependent change in the microwave tone of frequency $\omega_r$ transmitted through the resonator.\cite{Bianchetti2009} Since $E_J/E_C=39\gg 1$ and the anharmonicity is $\alpha_2/2\pi=-423\mhz$, our sample is operated well within the transmon regime. Charge dispersion is expected to amount to about $100\khz$ for $\omega_{01}$ and about $2.9\mhz$ for $\omega_{12}$. We have measured an energy relaxation time $T_1=0.84\us$, a phase coherence time $T_2^*=1.03\us$ and a spin-echo phase coherence time $T_2^{\mathrm{echo}}=1.11\us$.

We consider the Hamiltonian of a driven $n$-level qubit in a frame corotating at the drive frequency $\omega_{d}$,
\begin{align}
\label{eq:Hdriven}
H(\varphi)&=\hbar\sum_{j=0}^n (j\Delta+\alpha_j)\projection{j} \\
&\quad+\dfrac{1}{2}\hbar\Omega \sum_{j=0}^{n-1}(\sqrt{j+1}e^{-i\varphi}\ketbra{j+1}{j} +\mathrm{h.c.}), \nonumber
\end{align}
where $\alpha_j$ is the anharmonicity defined through the energy of the $j$th qubit energy level $\omega_{0j}=j\omega_{01}+\alpha_j$ (with $\alpha_0=\alpha_1=0$), $\Omega$ is the strength of the drive expressed in units of angular frequency and $\varphi$ its phase. $\Delta=\omega_{01}-\omega_d$ denotes the detuning between the frequency of the $\ket{0}\leftrightarrow\ket{1}$ transition and the drive frequency. The increase of the coupling strength $\propto\sqrt{j+1}$ follows from the off-diagonal matrix elements of the charge operator.\cite{Koch2007} In the two-level approximation, we restrict the Hamiltonian to the computational subspace of the qubit spanned by $\ket{0}$ and $\ket{1}$:
\begin{align}
H\approx\frac{\hbar}{2}(\Omega_x\sigma_x+\Omega_y\sigma_y+\Delta\sigma_z)
=\frac{\hbar}{2}\boldsymbol\sigma\cdot\mathbf{B},
\end{align}
with $\Omega_x=\Omega\cos\varphi$ and $\Omega_y=\Omega\sin\varphi$. This is precisely the Hamiltonian of a spin-half particle in an effective magnetic field $\mathbf{B}=(\Omega_x,\Omega_y,\Delta)$.\cite{Leek2007} By applying microwave frequency drives, we are able to experimentally control all three parameters of this effective field, which is our parameter space, to guide the qubit along a circular path $C$ with constant detuning $\Delta$. At first, the drive field is ramped up, tilting $\mathbf{B}$ so that it forms an angle $\vartheta=\arctan(\Omega/\Delta)$ with respect to the $z$-axis. Then, the phase of the drive is swept by $2\pi$, causing $\mathbf{B}$ to rotate once around the $z$-axis, either clockwise or anticlockwise. Finally, the drive is ramped back to zero. The time evolution of both quadratures of the drive is plotted in Fig.~\ref{fig:phase}(a). During this sequence, the solid angle subtended by the path as seen from the origin is $A=2\pi(1-\cos\vartheta)$. We then repeat the measurement for different driving strengths $\Omega$, thereby changing the solid angle $A$.

To determine the geometric phase experimentally, we employ an interferometric measurement [Fig.~\ref{fig:phase}(a)].
The leading and trailing resonant $\pi/2$-pulses implement a Ramsey measurement, while the resonant spin-echo $\pi$-pulse in the centre serves to cancel the dynamic phase.\cite{Jones2000,Leek2007} After preparing an equal superposition of the $\ket{0}$ and $\ket{1}$ states, the qubit traverses the path $C^-$ and the relative phase $2(\gamma_d-\gamma_g)$ acquired between $\ket{0}$ and $\ket{1}$ comprises both a dynamic ($\gamma_d$) and a geometric ($\gamma_g$) contribution. The spin-echo $\pi$-pulse then effectively flips the sign of the phase. As it traverses the second loop in the opposite direction, $C^+$, the qubit acquires the phase $2(\gamma_d+\gamma_g)$ since the dynamic phase, unlike the geometric phase, is independent of the direction of evolution. Thus, after following the contours $C^{-+}$, dynamic phase contributions cancel out and the qubit state has acquired a phase $\gamma=4\gamma_g$ which is purely geometric. Tracing out the contours in opposite direction, $C^{+-}$, simply inverts the sign of the phase, while following the contours twice in the same direction, $C^{++}$ or $C^{--}$, results in zero phase and serves as a control experiment [Fig.~\ref{fig:phase}(c)].

During the off-resonant pulse sequences $C^{\pm}$, the drive $\Omega$ is strong (corresponding to induced Rabi-frequencies $\lesssim110\mhz$) and therefore the higher levels of the qubit are populated as well. In order to visualise this population leakage, we consider the Hilbert space of a three-level system. Neglecting a global phase, any three-level state can be written as
$\ket{\psi}=
e^{\i\chi_1}\sin{\beta_1}\cos{\beta_2}\ket{0}
+e^{\i\chi_2}\sin{\beta_1}\sin{\beta_2}\ket{1}
+\cos{\beta_1}\ket{2}
$,
where $\chi_{1,2}\in[0,2\pi]$ are the phases of the ground and first excited state, respectively, relative to the phase of the second excited state, and $\beta_{1,2}\in[0,\pi/2]$ parametrise the populations. Therefore, every state can be represented as a point in the product manifold of a torus and an octant of a unit sphere.\cite{Arvind1997} Observing that $\beta_1\neq\pi/2$ while the ground state is subjected to the off-resonant drive $C^-$ [Fig.~\ref{fig:phase}(b)], we conclude that the instantaneous ground state leaves the computational subspace. The population of second excited state reaches up to $\approx 12\%$, showing the necessity to consider the higher levels in our experiment.
It is important to note that all resonant pulses effectively act in the subspace spanned by $\ket{0}$ and $\ket{1}$: in order to avoid exciting higher energy levels with the resonant pulses, we use an optimal control technique known as DRAG.\cite{Motzoi2009} Furthermore, to ensure that the second excited state $\ket{2}$ is depleted before the resonant pulses are applied, the off-resonant drive is adiabatically ramped down.

After qubit manipulation, which takes approximately $700\ns$, the population $P_z=(1-\sigma_z)/2$ of the first excited state is extracted by a dispersive readout.\cite{Bianchetti2009} The phase $\gamma$ the qubit has acquired during evolution is reconstructed with state tomography (ST). The second $\pi/2$-pulse of the Ramsey sequence rotates the qubit about either the $x$ or $y$ axis, and serves as tomography pulse.
In the absence of decoherence, the phase $\gamma$ is given by $\arctan\left(\eval{\sigma_y}/\eval{\sigma_x}\right)$ with $\eval{\sigma_x}=\cos\gamma$, $\eval{\sigma_y}=\sin\gamma$ and $\eval{\sigma_z}=0$. The same expression approximates $\gamma$ well even in the presence of decoherence, which reduces the size of the Bloch vector $(\sigma_x,\sigma_y,\sigma_z)$ to $0.47$ in our experiments, while keeping the ratio $\eval{\sigma_y}/\eval{\sigma_x}$ constant. Therefore, the geometric phase remains unaltered by the decoherence in our experimental setting.

In keeping with Berry's predictions for a two-level system, $\gamma=2A$,  we measure a phase $\gamma$ which is approximately twice the solid angle $A$ subtended by the path [Fig.~\ref{fig:phase}(c))]. However, the data clearly shows that $\gamma$ increasingly deviates from Berry's predictions as $A$ (and therefore also the drive $\Omega$) increases: the measured geometric phase is observed to be up to $15\%$ larger than expected.

These deviations can be explained by the presence of higher transmon levels. Defining the operator $N=\sum_{j=0}^n j\projection{j}$, the Hamiltonian in Eq.~(\ref{eq:Hdriven}) can be rewritten as $H(\varphi)=e^{-i\varphi N}H(0)e^{i\varphi N}$. It follows that, given an eigenvector $\ket{\Phi(0)}$ of $H(0)$, $\ket{\Phi(\varphi)}=e^{-i\varphi N}\ket{\Phi(0)}$ is an eigenvector of $H(\varphi)$. For the circular path $C$ described above, the geometric phase $\gamma_{\Phi(0)}$ acquired by the eigenvector $\ket{\Phi(0)}$ is then found to be~\cite{Berry84}
\begin{align}
\label{eq:BP}
\gamma_{\Phi(0)}
=i\int_0^{2\pi}\opelem{\Phi(\varphi)}{\dfrac{d}{d\varphi}}{\Phi(\varphi)}d\varphi
=2\pi\opelem{\Phi(0)}{N}{\Phi(0)}.
\end{align}
From Eq.~(\ref{eq:BP}), one indeed recovers the expression $\gamma_{\pm}^{(0)}=\pi(1\pm\cos\vartheta)$ for the geometric phase of a two-level system, where the sign $\pm$ corresponds to the positive and negative eigenvalue of $H(0)$, respectively. To compute the geometric phase for a multi-level system, we divide the Hamiltonian $H(0)=H_0+V$ into a block-diagonal part $H_0$ coupling the lowest two transmon levels, and a perturbative part $V$ coupling the remaining levels:
\begin{align}
\label{eq:Hsplit}
H_0&=\hbar\sum_{j=0}^n(j\Delta+\alpha_j)\projection{j}
    +\dfrac{\hbar\Omega}{2} (\ketbra{1}{0} +\mathrm{h.c.}),\nonumber \\
V  &=\dfrac{\hbar\Omega}{2} \sum_{j=1}^{n-1}(\sqrt{j+1}\ketbra{j+1}{j}
    +\mathrm{h.c.}). \nonumber
\end{align}
Substituting the expansion of the eigenvectors in V, $\ket{\Phi_j}=\ket{\Phi_j^{(0)}}+\ket{\Phi_j^{(1)}}+\ket{\Phi_j^{(2)}}+\ldots$, into Eq.~(\ref{eq:BP}), and retaining terms up to second order, the second excited state is found to contribute
\begin{eqnarray}
\label{eq:BPCorrection}
\Delta\gamma_{\pm}
   &=& \gamma_{\pm}-\gamma_{\pm}^{(0)} = \pi k\sin^2\vartheta \\ \nonumber
   & & \times \frac{2k(1\pm \cos\vartheta)+(2k\mp(3k+2)\cos\vartheta)\sin^2\vartheta}
   {(k\mp(3k+2)\cos\vartheta)^2}.
\end{eqnarray}
to the geometric phase $\gamma_{\pm}$, where $k\equiv\Delta/\alpha_2$. In the experiment we measure the quantity $\Delta\gamma \equiv 2(\Delta\gamma_{-}-\Delta\gamma_+)$, see Fig.~\ref{fig:phase}(c). We also note that $\Delta\gamma = 2\pi k \sin^4\vartheta/\cos\vartheta + \mathcal{O}(k^2)$ vanishes for large $\alpha_2$ as expected.

Since in the experiment $k\approx1/8$ is small and $|\Omega/\alpha_2|\ll1$, the expansion coefficients of $\ket{\Phi_j^{(1)}}$ and $\ket{\Phi_j^{(2)}}$ are small, and perturbative treatment is justified.
To verify the validity of the perturbative treatment, we simulated the qubit evolution for the pulse sequence in Fig.~\ref{fig:phase}(a), retaining four energy levels in a quantum master equation simulation, thereby taking into qubit population decay, decoherence and non-adiabatic effects arising from finite evolution time.\cite{Ai2009} Also, the Hamiltonian in Eq.~(\ref{eq:Hdriven}) with $n=4$ was numerically diagonalised to compute the geometric phase in the limit of perfect adiabaticity. We found that the perturbatively computed geometric phase  $\gamma$ using Eq.~(\ref{eq:BPCorrection}) differs from both simulations and numerical results by less than $2\%$.
\begin{figure}
  \centering
  \includegraphics[width=86mm]{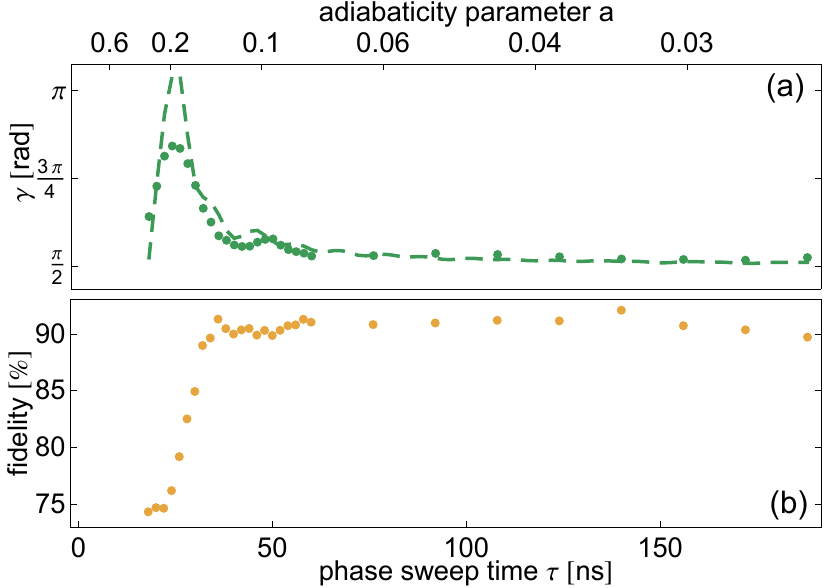}
  \caption{(Color online)
  (a) Measured phase $\gamma$ as function of phase sweep time $\tau$ for the solid angle $A=\pi/4$ at detuning $\Delta/2\pi=-45\mhz$. The dashed line is the phase obtained by numerically calculating the unitary time evolution of a four level transmon using the Schr\"odinger equation. The adiabaticity parameters $a$ for a given $\tau$ are indicated on the upper horizontal axis.
  (b) Fidelity of the geometric phase gates shown in (a) as a function of $\tau$.}
  \label{fig:adiab}
\end{figure}
Furthermore, we have measured $\gamma$ for a range of detunings $\Delta$ and have found good agreement with the geometric phase computed using perturbation theory [Fig.~\ref{fig:phase}(d)].

We have verified that the geometric phase does not depend on the dispersive coupling of the transmon to the resonator by tuning $\omega_{01}$ 
such that $\delta=\omega_r-\omega_{01}=1.58\ghz$ and comparing the data to the case shown in Fig.~\ref{fig:phase}(c), where $\delta=0.71\ghz$.

Finally, the transition from the adiabatic regime to the non-adiabatic regime was examined by measuring the acquired phase $\gamma$ at fixed solid angle $A=\pi/4$, changing only the phase sweep time $\tau$ [Fig.~\ref{fig:phase}(a)] but keeping the duration of the total sequence constant. The data in Fig.~\ref{fig:adiab}(a) shows that $\gamma$ is constant, i.e., independent of evolution time for $\tau$ larger than about $50\ns$, or equivalently, for adiabaticity parameters $a=\dot{\varphi}\sin(\vartheta)/|\mathbf{B}|\lesssim0.1$. The measured $\gamma$ also agrees well with the result obtained by numerically solving the Schr\"odinger equation. For values of $a>0.1$, when entering the non-adiabatic regime, 
we observe that $\gamma$ oscillates and varies by more than $50\%$. In this regime, $\gamma$ is a combination of dynamic and geometric phase because the spin-echo technique fails to cancel the dynamic phase: for non-adiabatic evolution, the state after the spin-echo $\pi$-pulse does not necessarily correspond to the initial state with $\ket{0}$ and $\ket{1}$ interchanged.

In the context of quantum information processing, the manipulation sequence could serve as a single qubit phase gate. Its performance can be assessed by computing the fidelity $F=\mathrm{tr}\sqrt{\rho^{1/2}\sigma\rho^{1/2}}$, where $\rho$ is the experimental density matrix processed with maximum likelihood~\cite{Jezek2003} and $\sigma$ is the expected density matrix for perfectly adiabatic evolution. We find that the fidelity of the gate averages $F=90\%$ in the adiabatic regime [Fig.~\ref{fig:adiab}(b)]. There, about $8\%$ of the loss in fidelity can be attributed to qubit decay, whereas inaccuracies in qubit preparation and qubit dephasing account for the remaining $2\%$. In the non-adiabatic regime a significant decrease in fidelity is observed, as expected.

In conclusion, we have measured the geometric phase in a multi-level system with small anharmonicity and observed that the two-level approximation breaks down for strong drives, as evidenced in our experiment by deviations of the geometric phase from the expected linear dependance on solid angle. We have modelled the contributions from the second excited state to the adiabatic geometric phase using time-independent perturbation theory.
By examining Berry's phase in the adiabatic limit and going to the non-adiabatic regime, we have shown that it is independent of evolution time for adiabaticity parameters $\lesssim0.1$. The phase in the non-adiabatic limit could potentially inherit some of the adiabatic phase's noise resilience, suggesting further experimental tests on how it is affected by noise in the control parameters.

This work was supported by the Swiss National Science Foundation
(SNF) and the EU project GEOMDISS.


\end{document}